# Giant non-volatile electric field control of proximity induced magnetism in the spin-orbit semimetal SrIrO$_3$

Arun Kumar Jaiswal[1], Robert Eder[1], Di Wang[2], Vanessa Wollersen[2], Matthieu Le Tacon[1], and Dirk Fuchs[1,#]

With its potential for drastically reduced operation power of information processing devices, electric field control of magnetism has generated huge research interest. Recently, novel perspectives offered by the inherently large spin-orbit coupling of 5*d* transition metals have emerged. Here, we demonstrate non-volatile electrical control of the proximity induced magnetism in SrIrO$_3$ based back-gated heterostructures. We report up to a 700% variation of the anomalous Hall conductivity $\sigma_{AHE}$ and Hall angle $\Theta_{AHE}$ as function of the applied gate voltage $V_g$. In contrast, the Curie temperature $T_C \approx 100$K and magnetic anisotropy of the system remain essentially unaffected by $V_g$ indicating a robust ferromagnetic state in SrIrO$_3$ which strongly hints to gating-induced changes of the anomalous Berry curvature. The electric-field induced ferroelectric-like state of SrTiO$_3$ enables non-volatile switching behavior of $\sigma_{AHE}$ and $\Theta_{AHE}$ below 60 K. The large tunability of this system, opens new avenues towards efficient electric-field manipulation of magnetism.

[1]Karlsruhe Institute of Technology, Institute for Quantum Materials and Technologies, 76021 Karlsruhe, Germany;

[2]Karlsruhe Institute of Technology, Institute of Nanotechnology and Karlsruhe Nano Micro Facility, 76021 Karlsruhe, Germany;

# 1. Introduction

Electric field (EF) control of magnetism in materials is of central importance for the development of sustainable and low power consumption information technology[1–3]. It is particularly challenging to achieve in ferromagnetic (FM) metals in which EF large enough to induce sizeable modifications of the magnetic state can generally not be applied. This has triggered intense research activity in particular on multiferroic oxides, where magnetic and ferroelectric order are inherently coupled, albeit generally weakly, or on heterostructures combining FM metals with a dielectric or ferroelectric gating material. There, the mechanism for the EF control of magnetism can be based on e.g. elastic strain mediation in combination with the inverse magnetostrictive effect, on voltage control of exchange coupling[1], or a modulation of the charge carrier density.

Latter approach is particularly relevant when magnetic phase, -moment, and -anisotropy depend noticeably on the density of states near the Fermi energy $E_F$ but is generally not very efficient in good FM metals. There, the EF-induced modulation of the carrier density $n$ is limited by the Thomas-Fermi screening length $\lambda_{TF} \propto (1/n)^{1/6}$ which only amounts to about 1 Å[4]. This has proven more effective in semiconductors[5] or in transition metal oxides (TMOs)[1] known for exhibiting significantly smaller $n$.

In heavier TMOs, such as the 4$d$ TMO $SrRuO_3$, electrostatic modulation of $n$ may not only result in changes of the magnetization and its anisotropy[6] but can also affect the integral of the Berry curvature (BC), thereby leading to changes of the anomalous Hall effect (AHE) - a fingerprint for the FM state in conductive materials[7]. AHE originates from the spin-orbit coupling (SOC) which is naturally present in heavy metals and enables, through an asymmetry in the scattering of spin-polarized electrons[8], an interplay between spin and charge on the electronic transport at the heart of the emerging field of spin-orbitronics.

SOC is particularly strong in the iridium-based 5$d$ TMOs of the Ruddlesden-Popper series $Sr_{n+1}Ir_nO_{3n+1}$ in which, however, in contrast to archetypical correlated 3$d$ TMOs, the electron-electron correlation strength is generally too small to host ferromagnetism. The iridates display SOC which is on a similar energy scale than that of the electron correlation or electronic bandwidth [9]. Therefore, they are at the verge of a magnetic ground state and may display AFM or FM properties as well, depending on the details of the Hubbard interaction $U$ and SOC[9]. The rather large Ir5$d$ and O2$p$ orbital hybridization in $SrIrO_3$ (SIO) ($n = \infty$) results in a semimetallic paramagnetic state[9–11]. We have recently shown that a FM state with large and positive AHE can be induced in SIO by proximity effect when putting it in direct contact with a FM insulator, $LaCoO_3$ (LCO)[12]. Recent first principle calculations on SIO/LCO heterostructures indicate that it originates from unconventional topology of the electronic band-structure of FM SIO[13].

In this work we demonstrate the EF control of magnetism in SIO heterostructures, evidenced through manifold increases of the anomalous Hall conductivity $\sigma_{AHE}$, the Hall angle $\Theta_{AHE}$ and the magnetoresistance. We further show that a non-volatile EF switching behavior is enabled by the EF-induced ferroelectric state of STO. The effects are discussed in terms of Rashba effect at the SIO/LCO interface and topological BC features of the SIO band structure, and we argue that the latter are more likely to account for the experimental observation.

## 2. Results and Discussion

### 2.1 Non-volatile electrostatic gating of SrIrO₃

Three terminal back-gating devices consisting of epitaxial SIO/LCO heterostructures were prepared by pulsed laser deposition and photolithography, see Methods and Supporting Information (SI). SIO and LCO film thicknesses of 10 monolayers demonstrate stable and reproducible proximity induced ferromagnetism in SIO. A scheme of the device layout is shown in **Figure 1**a. The high degree of crystallinity at the SIO/LCO interface is documented by the high-resolution scanning transmission electron microscopy (HR-STEM) micrograph in Fig. 1b. A gate-voltage $V_g > 0$ ($< 0$) usually results in an electron accumulation (electron depletion) in the SIO channel. The expected charge carrier modulation $\Delta n$ induced by electrostatic gating can be estimated by assuming a parallel-plate capacitor model to be of the order of 0.35 % (SI). The modulation thickness of the SIO channel is in principle limited by the Thomas-Fermi screening length[4], $\lambda_{TF}$, which can be significantly larger in semimetals than in good metals due to increased dielectric permittivity (SI). In addition, charge carrier localization may increase $\lambda_{TF}$ considerably as well. In LCO/SIO heterostructures, the first 6 SIO layers show insulating behavior [12]. Therefore, even for the rather short $\lambda_{TF} \approx 0.92$ ML as deduced from experiment, a distinct electric field can be expected at the SIO/LCO interface (SI).

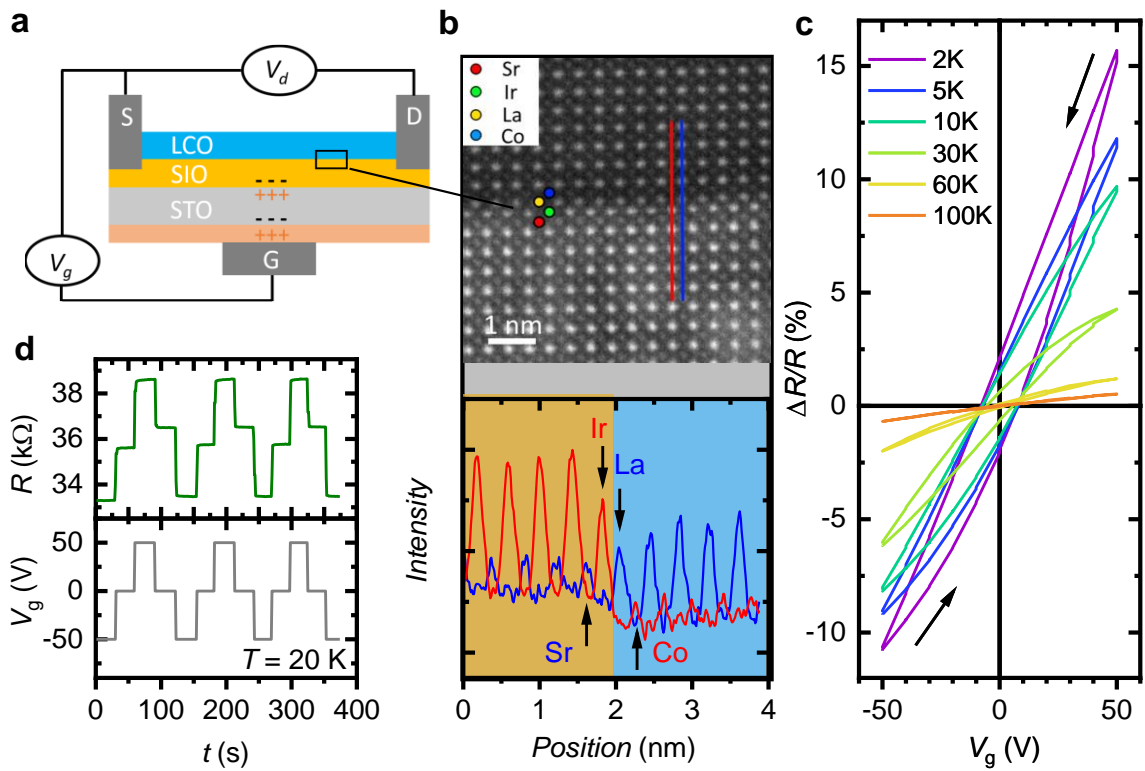

**Figure 1. Electrostatic gating of a SIO/LCO heterostructure. a** Scheme of the three terminal gating device structure. Source (S) and drain (D) contacts to the SIO/LCO interface were done by Al-wire bonding. The gate contact (G) was done by Pt-sputtering, silver paint and Al-wiring. As indicated, a positive gate voltage ($V_g$) usually results in a negative charging of the SIO channel. **b** (top) Cross-sectional HR-STEM micrograph of a typical SIO/LCO interface. Cations are indicated. The interface displays stoichiometric composition without distinct

structural defects. Element line -scans are indicated and shown below. (bottom) Element line-profiles generated from the cross-sectional HR-STEM. Scan direction (position) is from SIO to LCO. The red (blue) line verifies the atoms on the *B-* (*A-*) site of the *ABO$_3$*-perovskite structure, *i. e.*, Ir (Sr) and Co (La). Elements are indicated. **c** Variation of the channel resistance during sweeps of $V_g$ at different temperatures. Non-linear and non-volatile hysteretic behavior appears below about 60 K. Arrows indicated $V_g$-sweeping direction. **d** Dynamic switching behavior (top) for a specific $V_g$-sequence (bottom) at 20 K.

In Fig. 1c the relative change of the SIO resistance, $\Delta R/R$, is shown for $V_g$-sweeps at different temperatures *T*. $\Delta R/R$ increases with increasing $V_g$. The positive field coefficient ($dR/dV_g > 0$) suggests a hole-dominated conductivity in contrast to the generally observed negative Hall coefficient which hints to an electron-type transport[14–16]. This can be accounted for by the different mobilities of the electron and hole charge carriers demonstrated by magnetotransport[16] and electronic Raman scattering[17]. Note that electron- and hole-like pockets of semimetallic SIO sensitively depends on the structural properties (epitaxial strain)[11,14].

The increase of $\Delta R/R$ at any $V_g$ with decreasing *T* is consistent with the increase of the dielectric constant of STO, $\varepsilon_g$. We obtain $\Delta R/R \approx 15\%$ at $T = 2$ K for $V_g = 50$ V, which is larger than what is expected from simple electrostatics consideration and indicates decrease of electron mobility for electron accumulation in SIO/LCO (SI). Weak charge carrier localization also likely contributes to the monotonous increase of *R* with decreasing *T* [12], and explains the distinct increase of $\lambda_{TF}$.

At low temperatures, $\Delta R/R$ displays a non-linear and hysteretic behavior upon sweeping $V_g$, akin to gating effects reported in other STO heterostructures[18]. As such, the resistive state for $V_g = 0$ depends on the history (sweeping direction). The effect is only seen for $T \leq 60$ K, the temperature below which STO undergoes a phase transition to a ferroelectric-like state under electric fields $E \geq 2$ kV/cm[19,20]. The applied field strength ($E_g = +/- 5$ kV/cm) here is large enough to induce such transition and is therefore likely responsible for the observed and well-reproducible non-volatile switching behavior (Fig. 1d).

## 2.2 Electric field control of magnetism

We now turn to magnetotransport of the thin SIO layer. The insulating character of LCO confines electric transport solely to SIO and allows unambiguous selective characterization of the proximity induced FM state in SIO[12]. The AHE and the anisotropic magnetoresistance (*AMR*), two hallmarks of a FM metal, are related to the structure and ordering temperature of the FM state and are therefore useful quantities to characterize the magnetic properties of the material. The Hall resistivity $\rho_{xy}(\mu_0 H)$ and the magnetoresistance $MR = [\rho_{xx}(\mu_0 H) - \rho_{xx}^0]/\rho_{xx}^0$ ($\rho_{xx}^0 = \rho_{xx}(\mu_0 H = 0)$) with magnetic field $\mu_0 H$ applied perpendicular to the film surface are displayed in **Figure 2** for different $V_g$ at $T = 20$ K, *i. e.*, in the FM state of SIO ($T_C \approx 100$ K). For other temperatures, see SI.

The total Hall resistivity of SIO, $\rho_{xy}(\mu_0 H)$ can be decomposed in two components, respectively ordinary ($\rho_{OHE}$) and anomalous ($\rho_{AHE}$). $\rho_{OHE}$ is caused by Lorentz force and varies linearly with $\mu_0 H$ within the investigated field range. The large hysteresis seen in $\rho_{xy}$ (Fig. 2a) results from the anomalous contribution, typical for a FM metal[21]. In conventional magnetic systems, $\rho_{AHE}$ is proportional to the perpendicular magnetization $M$, $\rho_{AHE} = R^A \times M$, where $R^A$ depends on the longitudinal conductivity $\sigma_{xx}$[22]. On this basis, we can well describe our data using the empirical formula $\rho_{xy} = R^O \times H + R^A \times M$, where $M=(M_s \times \tanh(h \times (H \pm H_c)))$ is modelled using a modified Heaviside-step function. Here $M_s$, $h$ and $H_c$ are the saturation value, the slope at $H_c$ and the coercive field, respectively. $\rho_{AHE}$ is obtained after subtraction of the field-linear part contribution to $\rho_{xy}(\mu_0 H)$ and is shown in Fig. 2b (for $\rho_{OHE}$ see SI).

Remarkably, we observe a strong dependence of $\rho_{AHE}$ with $V_g$, from which we extract a relative increase of $M_s$: $(M_s(V_g)-M_s(0))/M_s(0)$ by a factor of 7 when going from $V_g = -50$ V to $+50$ V. A significant increase is also observed for $H_c$ and the saturation field $H_s$.

The MR obtained from longitudinal resistivity $\rho_{xx}(\mu_0 H)$ at 20K (see Fig. 2c) exhibits strong dependence with $V_g$ alike. It is also well described by the sum of two contributions, the classical Lorentz scattering ($MR \propto H^2$)[23] resulting in a positive contribution and spin-flip scattering ($MR \propto -M^2$), which contributes negatively to $MR$ in the FM state[24]. For $V_g = +50$ V $MR$ is dominated by the negative hysteretic contribution confirming strong FM character, in strong contrast to $V_g = -50$ V where $MR$ is bestridden by the classical positive contribution. For $\mu_0 H = 14$ T and $V_g = +(-)50$V $MR$ amounts to $-0.2\%$ ($0.4\%$).

As for the Hall effect, we can extract $M$ from the $MR$ data using a Heaviside step-function to model $M$ and fit the data, albeit only reliably for $V_g > 0$. The *effective magnetization M* perpendicular to the plane obtained this way is very similar to that obtained from AHE, as shown on a normalized scale in Fig. 2d. It is worth noting that this is not a trivial result as many different mechanisms – intrinsic (integral of the Berry curvature over occupied states [25]) as well as extrinsic (side-jump- and skew- impurity scattering[26]) can contribute to the AHE. For our SIO/LCO heterostructures the AHE was found to be intrinsic[12].

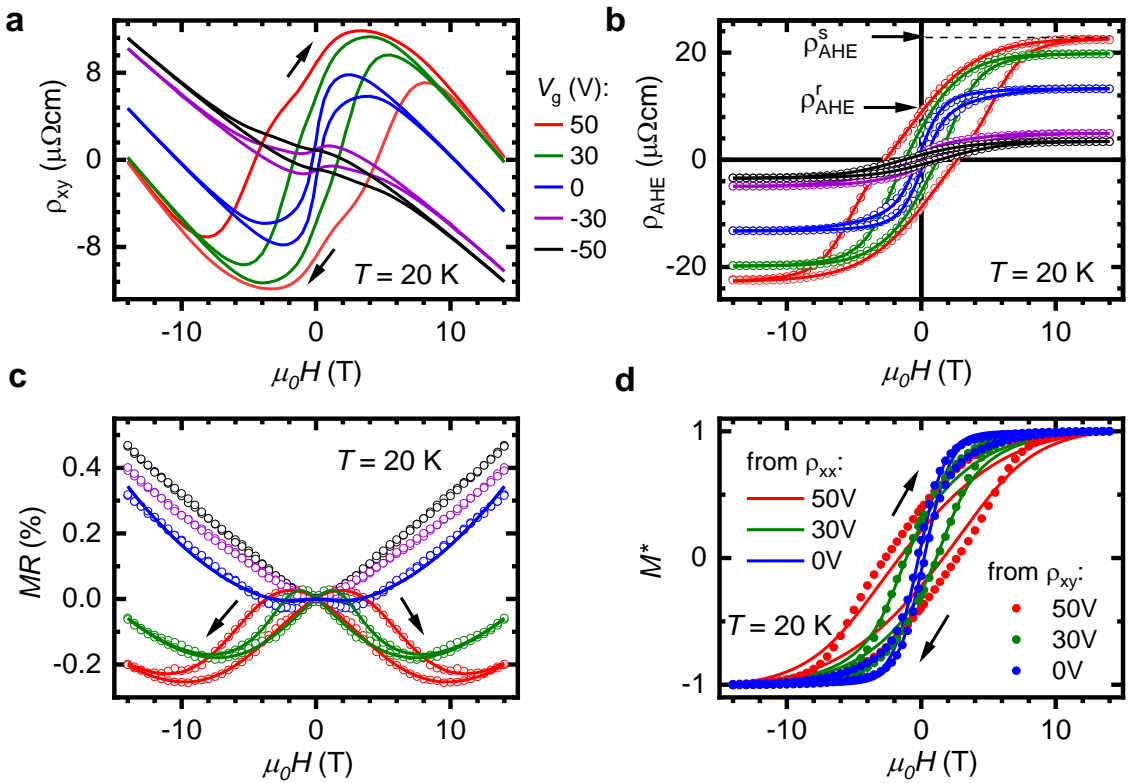

**Figure 2. Electric-field dependence of the magnetotransport. a** Hall resistivity $\rho_{xy}$ versus magnetic field $\mu_0 H$ for different $V_g$ at $T = 20$ K and **b** the extracted anomalous part of the Hall resistivity $\rho_{AHE}$ (symbols). Fits to the data (see text) are shown by solid lines. $\rho_{AHE}^s$ and $\rho_{AHE}^r$ define the saturated and remnant value of $\rho_{AHE}$. **c** The magnetoresistance $MR$ versus $\mu_0 H$ for different $V_g$ at $T = 20$ K and **d** the normalized perpendicular magnetization $M^* = M(\mu_0 H)/M_s$ as derived from the AHE and $MR$ versus $\mu_0 H$ for $V_g > 0$. Arrows indicate the field-sweep direction and color $V_g$.

Next, we take a closer look at the control of the FM state in SIO by electrostatic gating. In **Figures 3**a,b we show the saturated and the remnant anomalous Hall resistivity $\rho_{AHE}^s = \rho_{AHE}$ ($\mu_0 H = 14$T) and $\rho_{AHE}^r = \rho_{AHE}$ ($\mu_0 H = 0$), (see also Fig. 2b) versus $V_g$. The increase of $\rho_{AHE}^s$ with $V_g$ is non-linear and indicates the onset of a saturation for $|V_g| > 50$ V.

The field sweep shows hysteretic behavior due to the hysteretic gating effect of the ferroelectric-like STO, see Fig. 1c. As already mentioned, sweeping $V_g$ from -50 V to +50 V increases $\rho_{AHE}^s$ by a factor of 7, while $\rho_{AHE}^r$ increases by more than one order of magnitude. To the best of our knowledge, and in contrast to reports in *e.g.* 4d TMO SrRuO$_3$ and SrRuO$_3$/SIO heterostructures, the dependence of the *effective magnetization* with $V_g$ reported here is remarkably large[7,27,28]. Note that the unipolar electrostatic gating asymmetry of the effect for $\rho_{AHE}^r$ which is less affected for $V_g < 0$ (in contrast to $\rho_{AHE}^s$) than for $V_g > 0$ might be of practical interest for the realization of spintronic devices.

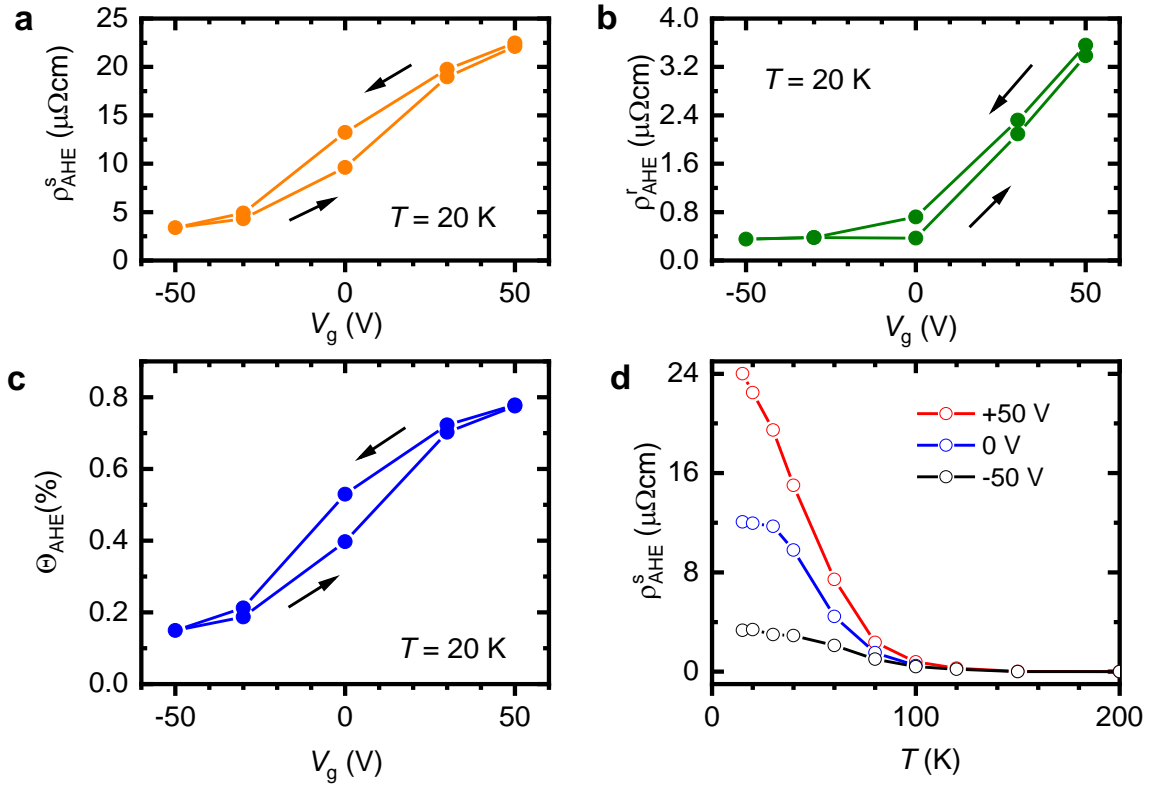

**Figure 3. Control of the ferromagnetic state by electrostatic gating. a** The saturated and **b** remnant anomalous Hall resistivity $\rho_{AHE}^s$ and $\rho_{AHE}^r$ versus $V_g$ at $T = 20$ K, **c** The Hall-angle $\Theta_{AHE} = \sigma_{AHE}^s/\sigma_{xx}^0$ versus $V_g$ at $T = 20$ K. Arrows indicate the field-sweep direction. **d** $\rho_{AHE}^s$ versus $T$ for different $V_g$.

Another quantity which is also highly relevant for spintronic purposes is the ratio between the saturated anomalous Hall conductivity $\sigma_{AHE}^S = \rho_{AHE}^S / [(\rho_{AHE}^S)^2 + (\rho_{xx}^0)^2]$ and the longitudinal conductivity $\sigma_{xx}^0 = \rho_{xx}^0 / [(\rho_{AHE}^S)^2 + (\rho_{xx}^0)^2]$ which defines the anomalous Hall angle $\theta_{AHE} = \sigma_{AHE}^S / \sigma_{xx}^0 = \rho_{AHE}^S / \rho_{xx}^0$. Generally, due to the large SOC, $\theta_{AHE}$ is much larger in the 5$d$ iridates compared to 3$d$ TMOs [22,29]. In Fig. 3c, $\theta_{AHE}$ is shown versus $V_g$ for $T = 20$ K. The gate voltage dependence is very similar to that of $\rho_{AHE}^S$, indicating a much stronger influence of $V_g$ on $\rho_{AHE}^S$ than on $\rho_{xx}^0$. This is in full agreement with the intrinsic nature of the AHE in SIO[12], where $\sigma_{AHE}$ does not depend on $\sigma_{xx}$. Sweeping $V_g$ from -50V to +50 V, increases $\theta_{AHE}$ by more than 500%. The temperature dependence of $M$ expressed by $\rho_{AHE}^S(T)$ is shown in Fig. 3d for different $V_g$. Although $\rho_{AHE}^S(T)$ is significantly enhanced for positive $V_g$, the onset of the effect $T_C \approx 100$ K is not much affected by $V_g$.

## 2.3 Anomalous magnetoresistance and magnetic anisotropy

Manipulation of the magnetic anisotropy or even switching of the magnetic easy-axis from in- to out-of-plane is of special practical interest. In the following, we discuss the EF-dependence of the magnetic anisotropy.

The proximity induced magnetism of SIO results in an angle-dependent anisotropic magnetoresistance $AMR(\alpha) = [\rho_{xx}(\alpha) - \rho_{xx}(90°)]/\rho_{xx}(90°)$ below $T_C$, where α is the angle between the in-plane magnetic field- and current-direction (Fig. 4b).

**Figure 4**a displays $AMR(\alpha)$ for different $V_g$ at 20 K and 14 T. Local maxima are seen close to 0, 90, 180 and 270 degrees, with amplitude that strongly depend on $V_g$. This can be described combining a two- and a four-fold angle-dependent component: $AMR(\alpha) = C_0 + C_2(\alpha) + C_4(\alpha)$, where $C_2(\alpha) = <C_2> \times \cos(2\alpha - \omega_2)$ and $C_4(\alpha) = <C_4> \times \cos(4\alpha - \omega_4)$. $C_0$, <$C_2$>, and <$C_4$> correspond to the amplitude of each component (offset angles ω$_2$ and ω$_4$ are allowed for each of them to account for backlash of the sample rotator).

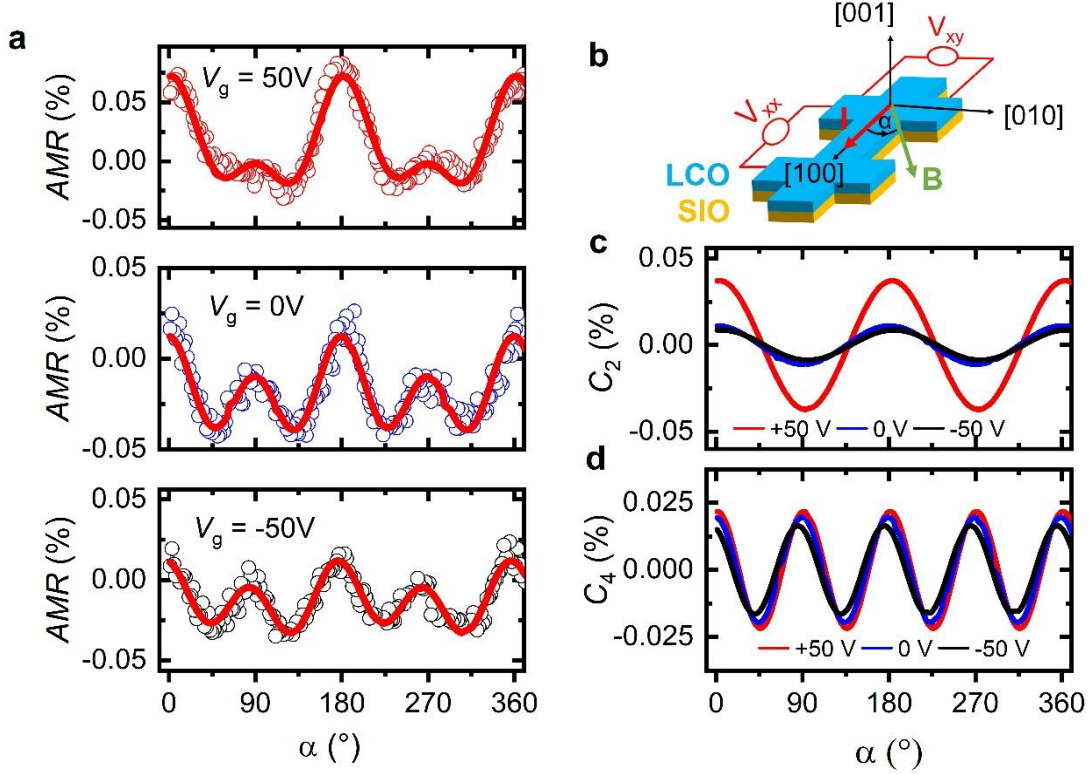

**Figure 4. Anomalous magnetoresistance of the SIO/LCO heterostructure. a** The angle-dependent anisotropic magnetoresistance $AMR(\alpha)$ for different $V_g$ at $T = 20$ K and 14 T. **b** Schematic of the pseudo-cubic crystallographic- and in-plane magnetic field direction with respect to the current flow direction. **c** The two-fold and **d** the four-fold component of $AMR(\alpha)$, shown in **a**. Fitting parameters are listed in SI.

The two-fold component shows maxima at $\alpha = 0°$ and $180°$, whereas the four-fold component displays minima at $\alpha = 45°+(n\times90°)$, $n = 0,1,2,$ and $3$. Fits to the data are shown by solid lines (for fitting parameters, see SI). The extracted two- and four-fold components of $AMR(\alpha)$ are shown in Figs. 4c and 4d, respectively.

The angle-dependence of the two-fold component is similar to that of the classical $AMR(\alpha)$[30], the spin-Hall magnetoresistance, $SMR(\alpha)$[31], or the spin-orbit magnetoresistance, $SOMR(\alpha)$[32]. However, the amplitude of ~ 0.04% is about two orders of magnitude larger compared to values generally reported for $SMR(\alpha)$ and $SOMR(\alpha)$[31,32] and more typical for the normal $AMR(\alpha)$, where the amplitude depends on the effective magnetization of the sample. This is confirmed by the strong increase of $<C_2>$ by a factor of 5 with increasing $V_g$, strongly reminiscent of the behavior of $\rho_{AHE}^S$ (so of $M$).

The four-fold component, on the other hand, does not depend on current direction and is thus related to the magnetocrystalline anisotropy. The minima positions correspond to reduced spin-flip scattering and indicate an in-plane <110> magnetic easy-axis direction[12]. In comparison, $<C_4>$ changes only by a factor of 1.6 when $V_g$ is increased which indicates slightly enhanced <110>-easy-axis behavior. The minima positions of the four-fold magnetocrystalline component are obviously not affected by $V_g$. The measurements suggest negligible influence of $V_g$ on the symmetry or strength of the magnetocrystalline anisotropy.

## 2.4 Discussion

We have presented a series of experimental results demonstrating efficient EF-control of the magnetic behavior of SIO-based heterostructures. Very large changes of $\rho_{AHE}$ (×7) and $\Theta_{AHE}$ (×5) when sweeping the gate voltage from $V_g$ =-50V to +50 V are observed, with a strong asymmetry of the effect, which is essentially occurring for $V_g > 0$. In contrast, the magnetic anisotropy and $T_C$ are rather unaffected. Note that in a Stoner description of SIO ferromagnetism, one could expect $T_C$ to be sensitive to changes of the charge carrier density $\Delta n$ upon gating. Given the weakness of the electrostatic modulation ($\Delta n/n \approx 0.35\%$, see SI) in this system, the variation of $T_C$ might simply be too small to be detected.

How to understand in this context the huge EF-induced changes of the AHE? Two natural options are the EF-dependence of the Rashba effect at the SIO/LCO interface[33–35] [33–36] on the one hand and topological features of SIO band structure on the other hand.

The strength of the Rashba effect can be controlled by an applied strong EF through the linear dependence of the Rashba coefficient $\alpha_R$ on $E$ for free charge carriers[37]. In order to obtain significant effects, large electric field strengths of the order of V/nm have to be applied[33]. Such a large EF may also result in a strong coupling to electronic structure via orbital deformation and anomalous band splitting[38]. The resulting momentum dependent 'Rashba equivalent' magnetic field is expected to affect $\sigma_{AHE}$ quadratically (so symmetrically) on EF or $V_g$[33], in strong contrast with our experimental observation, ruling out Rashba origin as the main source for the EF tunability of the AHE reported here. Note, electric field strength applied here is only of the order $10^{-4}$ V/nm.

Topological band properties of SIO may also contribute to the AHE. The scattering-independent intrinsic contribution to the AHE comes from the Berry phase supported anomalous velocity. An interesting aspect of the intrinsic contribution to the AHE is that the Hall conductivity $\sigma_{AHE}$ is given as an integral of the BC, $\Omega^z$, over all occupied states below the Fermi energy[39]:

$$\sigma_{AHE} = -\frac{e^2}{\hbar} \int \frac{d\mathbf{k}}{(2\pi)^3} f(\epsilon_k) \Omega^z \quad (1)$$

Here, $f(\varepsilon_k)$ is the Fermi-Dirac distribution function.

A direct connection between BC and magnetotransport has been recently reported for topological insulator $Bi_2Se_3$[40] in which gating-induced upshift of $E_F$ was found to increase the contribution of conduction electrons to BC and to increase the spin Hall conductivity. In magnetic oxides with complex band structures, the intrinsic mechanism for the AHE and the spin Hall effect (SHE) is the same[22] and depends on the detailed properties of the momentum-space BC. The presence of band crossing points close to $E_F$ can affect the BC and even result in a sign-change of the AHE[41].

For SIO, the large intrinsic SHE previously reported[42,43] supports the existence of BC anomalies. Furthermore, recent first principle calculations on SIO/LCO heterostructures unraveled a ferromagnetic band structure for the tetragonal structured SIO exhibiting non-trivial topological features (double Weyl points above and below $E_F$) responsible for a large AHE[13]. Their location in the Brillouin zone of the tetragonal distorted FM SIO is shown in Fig. 5b.

They indeed contribute positively to the AHE and the integral BC results in $\sigma_{AHE} = 7.5\ \Omega^{-1}\text{cm}^{-1}$, well comparable to the experimental value of $3\ \Omega^{-1}\text{cm}^{-1}$ for SIO/LCO below 30 K[12].

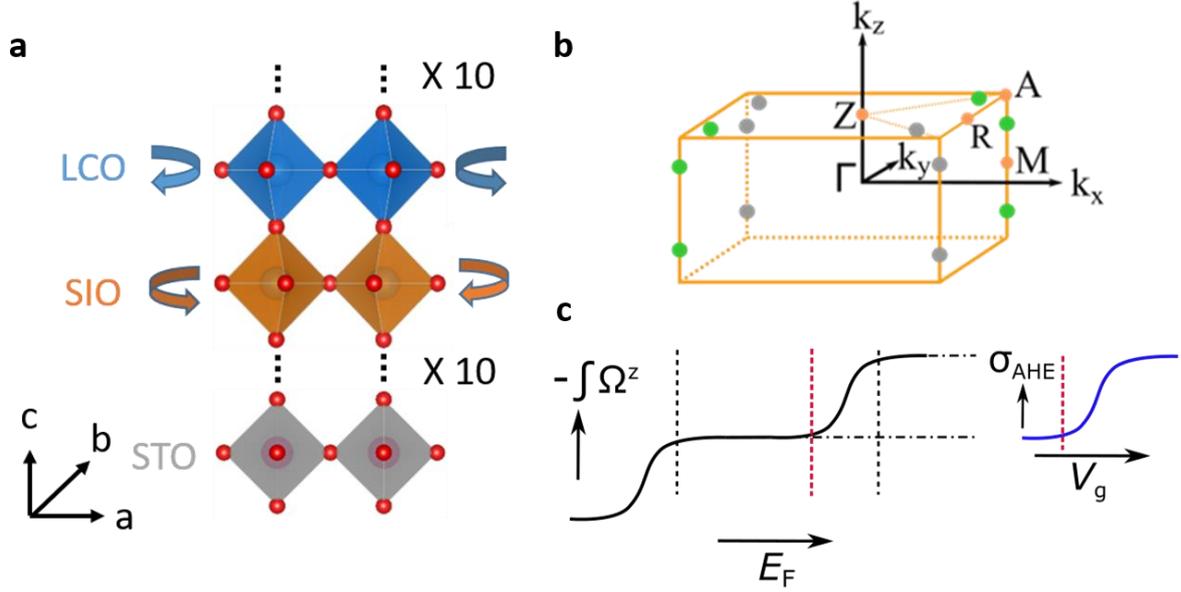

**Figure 5. Structure and topology of FM SIO and influence on the AHE. a** Structure of the SIO/LCO heterostructure. Only the pseudo-cubic perovskite cells are indicated. In-plane lattice parameters are the same as for STO substrate. Out-of-plane lattice parameter of SIO (LCO) is larger (smaller) compared to that of STO. The ($a^0a^0c^-$) octahedral rotation pattern of LCO and SIO is indicated by arrows. **b** The Brillouin zone of the tetragonal distorted FM SIO. High symmetric points are indicated. Weyl points with positive (green) and negative (grey) chirality are highlighted. Data were taken from Ref.[13]. **c** Integral of the BC as a function of $E_F$. Weyl crossing points occurring 20 meV below $E_F$ (indicated by black dashed line) result in a strong increase of $-\int \Omega^z$. Similar behavior is expected when $E_F$ (indicated by red dashed line) is further increased above the second double of Weyl points located 10 meV above. An increase of $V_g$ and $n$ also increases $E_F$ resulting in a similar rise of $\sigma_{AHE}$ compared to that of $-\int \Omega^z$ which is shown on the right.

This allows us to sketch a scenario for the EF control of magnetism in SIO reported here: since double Weyl points are located only 10 meV above $E_F$, an EF-induced charge carrier accumulation ($V_g > 0$) may shift $E_F$ across those band-crossing points (indicated by the dashed black lines in Fig. 5c). The shift of $E_F$ can be estimated by the electron doping $\Delta n = 3.5 \times 10^{18}$ cm$^{-3}$ for $V_g = +50$V (see SI) and the density of states (DOS) at $E_F$ (see SI of Ref.[13]). Considering the integrated density of states, a shift of 0.01 eV by $\Delta n$ and therefore access of the second Weyl point by electric gating is very likely. When this occurs, the integral BC increases abruptly and so do $\sigma_{AHE}$ or $\Theta_{AHE}$[44,45]. The minor influence of $V_g$ on the magnetic ordering temperature $T_C$ and anisotropy tend to favor such topology-based scenario [45].

Many properties such as the orbital magnetization and anomalous Hall conductivity can be expressed in terms of Berry phases, connections, and curvatures and are therefore directly related to each other. Hence, $\rho_{xx}(\mu_0 H)$ and $MR$ will naturally be affected by the AHE alike[46]. As such, the direct relation between $\rho_{AHE}(\mu_0 H)$ and $MR$ emphasized earlier (Fig. 2d) also favors the topological scenario.

## 3. Conclusion

In summary, we have shown that electrostatic gating allows for a very large tunability of the proximity induced $\sigma_{AHE}$, $\Theta_{AHE}$ and *MR* in SIO/LCO heterostructures likely rooted by the singular band structure of SIO. The results demonstrate that the magnetic properties of FM topological materials can be very effectively controlled by electric fields, offering a promising avenue for realizing energy efficient spintronic devices.

## 4. Experimental Section/Methods

*Sample preparation*

The SIO/LCO heterostructures were grown on (001) oriented SrTiO$_3$ (STO) substrates by pulsed laser deposition. First, 10 MLs of SIO were deposited on TiO$_2$-terminated (001)-oriented STO substrate, followed by the deposition of 10 MLs of LCO. Film-thickness and layer-by-layer growth were controlled in-situ by reflection high energy electron diffraction (RHEED). More details on film preparation are described elsewhere[12,47]. After film preparation, microbridges (40 μm width and 200 μm length) were patterned by standard ultraviolet photolithography and Ar-ion etching. Next, the STO substrate was thinned down from the backside to about 0.1 mm to increase possible electric field strength. The back-gate electrode was provided by Pt-sputtering, silver-paste and Al-wiring, whereas source and drain contacts to the SIO/LCO interface were done by ultrasonic Al-wire bonding.

The structural properties of the SIO/LCO heterostructures were analyzed by x-ray diffraction using a Bruker D8 DaVinci diffractometer and high-resolution transmission electron microscopy (HRTEM) as shown elsewhere[12].

*Electronic Transport and data analysis*

Measurements of the electronic transport were carried out using a physical properties measurement system (PPMS) from Quantum Design. The modulated (2Hz) source-drain sample current was typically 1-10 μA. A sample rotator HR- 133 was used for angle-dependent magnetoresistance measurements. The gate-voltage was provided externally by a Keithley 6517B Electrometer, constrained to 50V by the PPMS. Before the measurements, the sample were kept for 12 hours in the cryostat at 2K to stabilize sample and the gate-voltage was ramped up and down several times. *AMR*($\alpha$) measurements were done starting from $V_g$ = -50 V to $V_g$ = +50 V.

The longitudinal and transversal resistivity were symmetrized and anti-symmetrized with respect to magnetic field to obtain $\rho_{xx}(\mu_0 H)$ and $\rho_{xy}(\mu_0 H)$, respectively. $\rho_{AHE}(\mu_0 H)$ was deduced from $\rho_{xy}(\mu_0 H)$ by subtraction of the linear part, *i. e.*, $\rho_{OHE}(\mu_0 H)$. Angle-dependent $\rho_{xx}(\alpha)$ was corrected with respect to sample wobbling and offset resistance. All the fitting routines described in the text were carried out with MATLAB and fitting parameters are listed in the SI.

**Supporting Information**

Supporting Information is available from the Wiley Online Library or from the author.

**Acknowledgements**

A. K. J. acknowledges financial support from the European Union's Framework Programme for Research and Innovation, Horizon 2020, under the Marie Skłodowska-Curie grant agreement No. 847471 (QUSTEC). D. F. thanks S. Rom and T. Saha-Dasgupta from the S. N. Bose National Centre for Basic Sciences (Kolkata, India) for fruitful discussions. We are grateful to R. Thelen and the Karlsruhe Nano-Micro Facility (KNMF) for technical support.

# Supplementary Information

# Giant non-volatile electric field control of proximity induced magnetism in the spin-orbit semimetal SrIrO$_3$

A. K. Jaiswal, R. Eder, D. Wang, V. Wollersen, M. Le Tacon, and D. Fuchs[#]

**Device preparation.** Three terminal back-gating devices comprising epitaxial SIO/LCO heterostructures on (001) oriented SrTiO$_3$ (STO) substrates were prepared by pulsed laser deposition and standard photolithography. After film preparation, microbridges (40 μm width and 200 μm length) were patterned by standard ultraviolet photolithography and Ar-ion etching. Next, the STO substrate was thinned down from the backside to about 0.1 mm to increase possible electric field strength. The back-gate electrode was provided by Pt-sputtering, silver-paste and Al-wiring whereas source and drain contacts to the SIO/LCO interface were done by ultrasonic Al-wire bonding. The final device is shown in Fig. S1.

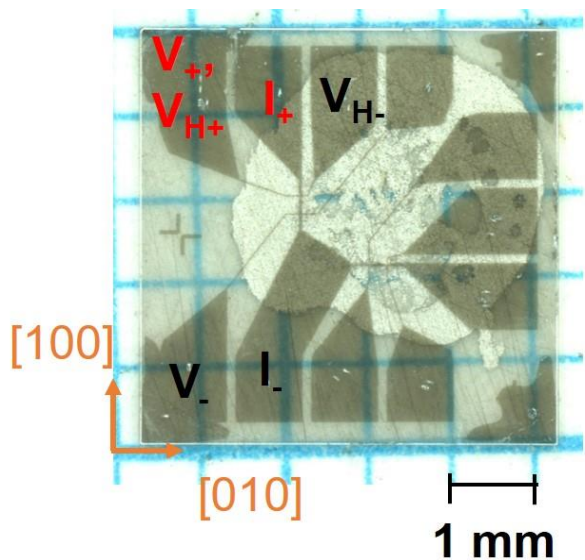

**Figure S1. Three terminal back-gating device.** Top view of the device. Contact pads for current *I*, voltage *V* and Hall-voltage *V*$_H$ are indicated. Microbridges (40μm×200 μm) along the [100]$_{pc}$ and [010]$_{pc}$ pseudo-cubic direction of SIO are available. Al-bonds are located at the end of the contact pads. Back side of the thinned STO substrate is covered with silver paste for contacting the gate. Scale is visualized by millimeter paper.

**Electric field-effect in SIO heterostructures.** In the following, we show electric field-effect experiments on two different types of SIO heterostructures, a STO/SIO/STO/LCO (sample A) and a STO/SIO/LCO heterostructure (sample B). Both samples display a layer thickness of 10 monolayer (ML) of SIO and LCO. In contrast to sample B, where SIO shows proximity induced magnetism at the SIO/LCO interface, SIO is not magnetic in sample A, where proximity effect is suppressed by a 4 ML thick STO insertion layer. The field effect devices are prepared as documented in the main text. From a back-gate capacitor model, the change in the charge carrier concentration $n$ at $V_g = +50$ V amounts to $\Delta n = 3.5\times10^{18}$ cm$^{-3}$, see below. For sample A ($n = 1.9\times10^{21}$ cm$^{-3}$) and B ($n = 1\times10^{21}$ cm$^{-3}$) this corresponds to a relative change $\Delta n/n$ of 0.18% and 0.35%, respectively.

The expected charge carrier modulation $\Delta n$ induced by electrostatic gating was estimated by assuming a parallel-plate capacitor model:

$\Delta n = \frac{\varepsilon_0 \varepsilon_g E_g}{e d_{ch}}$, where $\varepsilon_0$ and $\varepsilon_g$ are the permittivity of vacuum and STO, $E_g$ the gate electric field strength, $e$ the elementary charge and $d_{ch}$ the channel thickness. Due to the $T$-dependence of $\varepsilon_g$ the gating effect becomes increasingly efficient with decreasing $T$. Assuming $\varepsilon_g \approx 5000$ for $T <$ 40 K and $E_g = 5$ kV/cm ($V_g \approx 50$ V), $\Delta n$ amounts to $3.5\times10^{18}$ cm$^{-3}$. With respect to the charge carrier density $n \approx 10^{21}$ cm$^{-3}$ of the SIO channel this corresponds to a charge carrier modulation of about 0.35 %.

To probe charge carrier concentration, Hall measurements were carried out. The transversal Hall resistance $R_{xy}$ of SIO films typically show linear behavior for magnetic fields $B \leq 14$ T, despite the semimetallic behavior of SIO. Considering two types of charge carriers, *i. e.*, electrons and holes, a two-band model is used to describe $R_{xy}$:

$$R_{xy} = \frac{(n_h\mu_h^2 - n_e\mu_e^2) + \mu_h^2\mu_e^2(n_h - n_e)B^2}{(n_h\mu_h + n_e\mu_e)^2 + \mu_h^2\mu_e^2(n_h - n_e)^2 B^2} \qquad (S1)$$

which in the low field limit results in:

$R_{xy}/B \propto (n_h\mu_h^2 - n_e\mu_e^2)$ (S2)

where $n_h$, $n_e$, and $\mu_h$, $\mu_e$ are the density and mobility of holes and electrons, respectively. In Fig. S2(a), the relative change of the SIO channel resistance of sample A is shown versus the gate voltage $V_g$ for different $T$. The hysteretic behavior for $T < 60$ K is caused by electric-field induced ferroelectric -like state in STO, see also main text. Interestingly, $R$ increases for $V_g >$ 0. Generally, $V_g > 0$ results in an electron accumulation, *i. e.*, in an increase of $n_e$ ($\Delta n_e > 0$). Therefore, a decrease of the electron mobility is expected where:

$\Delta n_e < -\Delta(\mu_e^2)$ (S3)

A change of $n_h$ and $\mu_h$ is neglected for $V_g > 0$. With respect to Eq. (S3), a decrease of $\mu$ with increasing $n$ is usually observed for SIO and oxide heterostructures at low $T$ for $n > 10^{19}$ cm$^{-3}$.

For sample A, $\Delta R/R$ amounts to about +0.5% (-0.5%) for $V_g = 50$ V ($V_g = -50$V) and $T \leq 30$ K. Assuming $R \propto (n\mu)^{-1}$ the relative change of $R$ is given by:

$\Delta R/R = - (\Delta n/n + \Delta\mu/\mu)$ (S4)

Therefore, for $\Delta n/n = 0.18%$, the relative decrease of $\mu$ should be around $\Delta\mu/\mu = -0.68%$, which is well in the range of what has been observed for SIO (see Manca *et al.*, Phys. Rev. B. 97 (2018)).

The $T$-dependence of the gating effect is shown in Fig. S2(b), where $\Delta R/R$ versus $T$ is plotted for $V_g = +50$ V. As expected, the $T$-dependence is very similar to that of the permittivity $\varepsilon(T)$ of the STO gate material, *i. e.*, a distinct increase below 100 K and saturation below about 30 K for $E > 1.5$ kV/cm.[3]

From these observations, the gating effect on the STO/SIO/STO/LCO heterostructure (sample A) is well described by simple electrostatics and the semimetallic behavior of SIO. However, for the STO/SIO/LCO heterostructure (sample B), where LCO is in direct contact with SIO, the gating effect is different. The relative change $\Delta R/R$ for $V_g = +50$ V amounts to 15% at 2K and is about 30 times larger compared to sample A even though the field-induced electron accumulation should be the same in both samples. With respect to Eq. (S4) a decrease of $\Delta\mu/\mu \approx -15.35\%$ is expected, which may hint towards weak charge carrier localization in SIO. This is well consistent with the $R$ versus $T$ behavior of both samples (see SI of Ref. [12]). The $T$-dependence of $\Delta R/R$, see Fig. S2(d), also differs significantly to that of sample A. $\Delta R/R$ steadily increases with decreasing $T$ down to 2K.

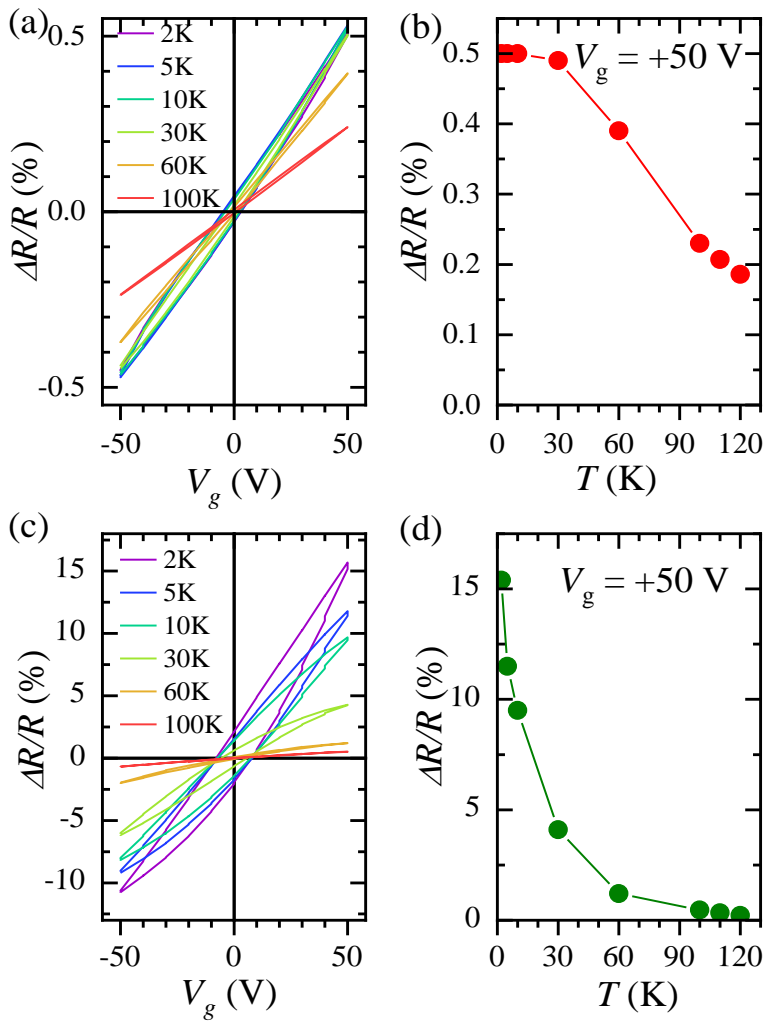

**Figure S2. Electric field-effect in SIO heterostructures.** The electric field-effect on a STO/SIO/STO/LCO (sample A, see (a) and (b)) and a STO/SIO/LCO heterostructure (sample B, see (c) and (d)). Both samples display a layer thickness of 10 monolayer (ML) of SIO and LCO. In contrast to sample B, where SIO shows proximity induced magnetism at the SIO/LCO interface, SIO is not magnetic in sample A, where proximity effect is prevented by a 4 ML thick STO insertion layer. The relative change of the SIO channel resistance is shown versus the gate voltage

for different *T* in (a) and (c). The *T*-dependence of the relative change of the SIO channel resistance $\Delta R/R = [R(V_g)-R(0)]/R(0)$ is displayed in (b) and (d).

**Electric screening.** The effective modulation thickness of the SIO channel is in principle limited by the Thomas-Fermi screening length $\lambda_{TF}$, which describes the penetration of the electric field *E* into the channel material as a function of distance *z* from the gate-channel interface: $E(x) = E(0)\exp(-z/\lambda_{TF})$, where $\lambda_{TF} = (\varepsilon_0\varepsilon_k/e^2N(E_F))^{1/2}$. Here, $\varepsilon_k$ and $N(E_F)$ are the dielectric permittivity and density of states near the Fermi energy $E_F$ of the channel material. In the limit of a 3-dimensional electron gas this results in $\lambda_{TF} \approx (\varepsilon_0\varepsilon_k\hbar^2\pi/me^2)^{1/2} \times (\pi/(3n))^{1/6}$, with the particle mass *m* and density *n*. Assuming $\varepsilon_k = 10$ (typical for semimetals – large values of $\varepsilon_k = 30$ have been reported for $Sr_2IrO_4$[48,49] and even higher values $\varepsilon_k \approx 600$ for other iridates [S1]) $\lambda_{TF}$ amounts to about 6 Å, which is significantly larger compared to good metals ($\lambda_{TF} \sim 1$ Å), however still smaller compared to the SIO film thickness of 10 ML ($\approx$ 40 Å). To increase $\lambda_{TF}$ by one order of magnitude ($\lambda_{TF} \geq 60$ Å), $\varepsilon_k$ has to be increased by two orders of magnitude ($\varepsilon_k \geq 1000$).

On the other side, in case of electron localization classical Lindhard theory and derived formula for Thomas Fermi screening length $\lambda_{TF}$, which assumes uniform electron gas, does not hold any more. Localized electrons cannot move freely, and hence electric fields are not fully screened. In this case, even weak charge carrier localization such as likely present in sample B will result in a significant drop of the dynamic screening and a much larger screening length compared to the expected value of $\lambda_{TF}$. We want to point out here, that the first few layers of SIO on STO do show insulating behavior. As documented in Fig. 2b of Ref. [12], the first 6 layers of SIO in the SIO/LCO heterostructures are obviously not conductive. Therefore, distinct damping of the electric field in SIO may start not before the 7th layer!

To probe the thickness dependence of the electric field effect in more detail, we have prepared a set of LCO(10 ML)/SIO(*x*ML) heterostructures with different SIO film thickness (*x* = 6 – 15 ML). In Fig. S3(a) we have shown the normalized resistance versus *T* for different *x*. Obviously, for *x* < 10 ML the resistance strongly increases with decreasing *T*. For *x* = 6, the films are insulating below 100K. The gating effect, *i. e.*, the relative change of the resistance with gate voltage $\Delta R/R = [R(+50V)-R(0)]/R(0)$ versus *x* is displayed in Fig. S3(b) for *T* = 10K. For x > 10 ML the gating effect drops down exponentially, as expected from classical electric field damping. Assuming charge carrier modulation $\Delta n/n$ proportional to $\Delta R/R$ and proportional to electric field strength *E* we can deduce electric field damping. From the exponential drop, $\lambda_{TF} \approx 0.92$ ML is deduced, which compares well with the estimated value of 6 Å. This result is also consistent with the assumption that the first 6 SIO layers are insulating. Figure S4 demonstrates $E(x)$ assuming insulating behavior for the first 6 ML of SIO and $\lambda_{TF} = 0.92$ ML. A distinct electric field can be expected in the magnetic active SIO layers. As already shown in Ref.[12], the magnetic proximity effect is triggered by interfacial charge transfer from Ir4+ to Co2+ and limited to the first or second SIO layer close to the SIO/LCO interface.

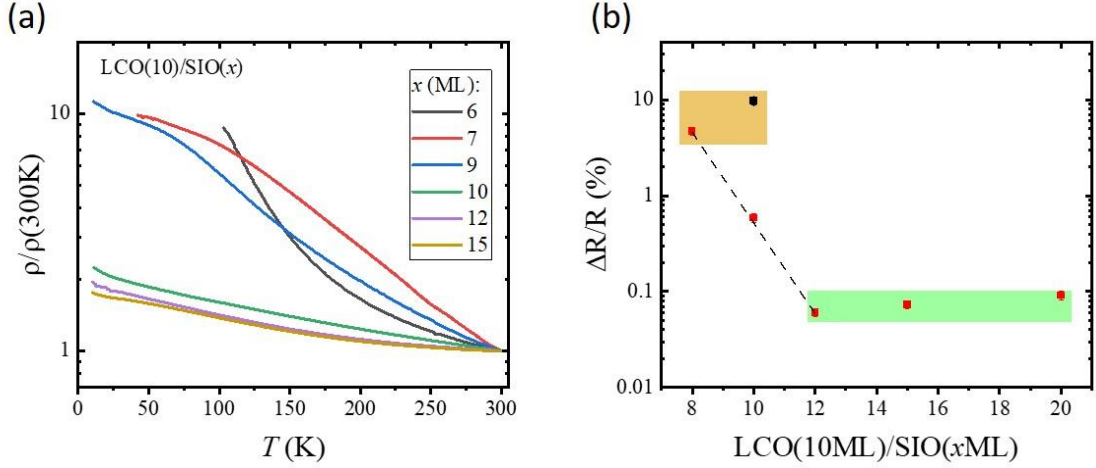

Figure S3. (a) Normalized resistivity of LCO/SIO heterostructures for different SIO layer thickness $x$ as a function of $T$. For $x \leq 6$ ML resistance is above measurement limit below 100K. (b) Relative change of the resistance with gating voltage $V_g$ as a function of SIO layer thickness $x$ at $T = 10$K. $\Delta R/R = [R(+50V)-R(0)]/R(0)$. Red symbols correspond to samples with a STO-gate thickness of 0.17 mm. The black symbol corresponds to the heterostructure sample (B) – see also main text-, where the STO thickness was only 0.1 mm.

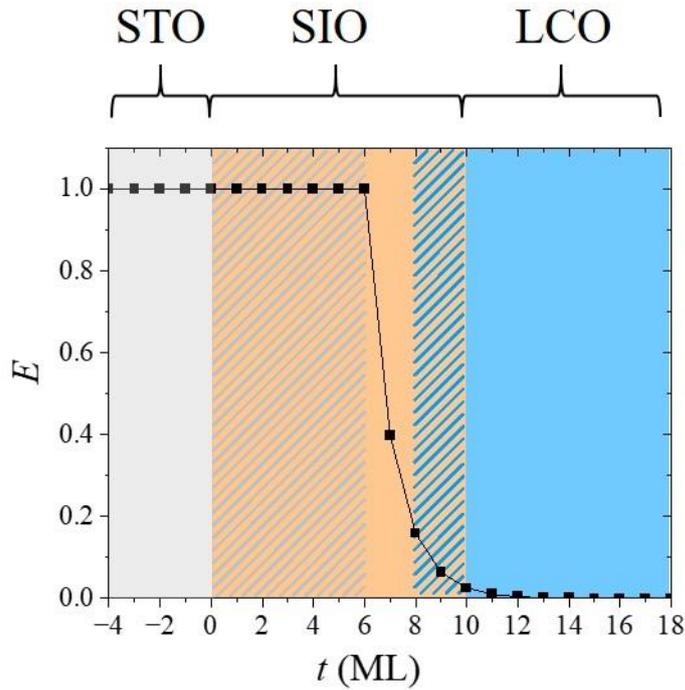

Figure S4. Screening of the electric field in the LCO/SIO/STO heterostructures. The insulating layers and magnetic active layers in SIO are indicated by grey and blue shaded region. The screening length was chosen to be 0.92 ML as deduced from experiment (see above). Distinct residual field is expected at the SIO/LCO interface.

**Field-induced ferroelectric-like sate in SrTiO$_3$.**

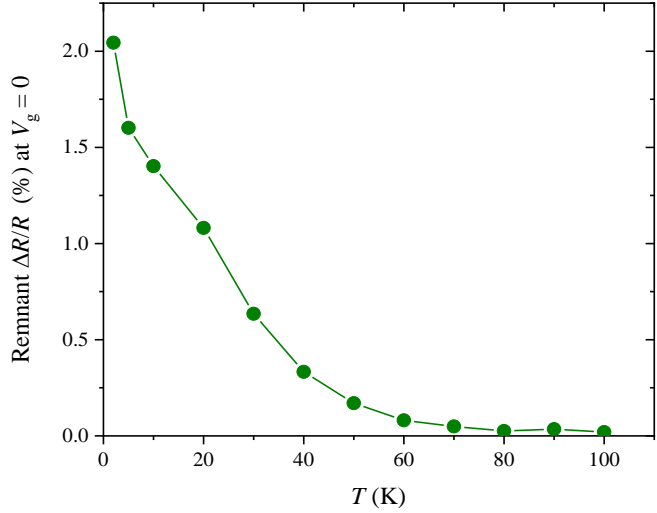

**Figure S5. Remnant resistivity change versus $T$.** The remnant resistivity change $\Delta R/R$ at $V_g = 0$ as a function of temperature. The resistivity change was averaged for decreasing and increasing $V_g$ to 0. The onset of $\Delta R/R$ indicates field-induced ferroelectric-like state below 60 K.

**Magnetotransport of SIO/LCO heterostructures.** The normal magnetoresistance $MR = (\rho_{xx}(\mu_0 H) - \rho_{xx}(0))/\rho_{xx}(0)$ and the Hall resistivity $\rho_{xy}(\mu_0 H)$ with magnetic field $\mu_0 H$ perpendicular to the film surface are displayed in Fig. S6 for different $V_g$ (+50, 0, and -50 V) at various temperatures for $T < T_C$ ($T = 20, 30, 40$, and 80 K). Magnetotransport reflects the FM state of SIO ($T_C \approx 100$ K). $MR$ (see Fig. S6a) is well described by the sum of two contributions, the classical Lorentz scattering ($MR \propto H^2$) resulting in a positive contribution to $MR$ and spin-flip scattering ($MR \propto -M^2$), which is effectively suppressed in the FM state and leading to a negative contribution to $MR$ below $T_C$. The Hall resistivity $\rho_{xy}(\mu_0 H)$ (see Fig. S6b) is also best described by the sum of two components, *i. e.*, the ordinary Hall resistivity $\rho_{OHE}$ caused by Lorentz force which for SIO films is usually found to be linear to $\mu_0 H$ in that field-range, and a hysteretic anomalous part $\rho_{AHE}$, typical for a FM metal. $\rho_{AHE}$ (see Fig. S6c) has been obtained by subtracting the linear contribution $\rho_{OHE}$ (see Fig. S6d) from $\rho_{xy}(\mu_0 H)$. Magnetization obviously increases with decreasing $T$ and positive gate voltage ($V_g > 0$) resulting in a continuous increase of $\rho_{AHE}$ and the negative contribution to $MR$.

The slope of $\rho_{OHE}$ versus $T$ is always negative indicating electron-like transport. However, slope of $\rho_{OHE}$ increases with increase $V_g$. Within a two-type carrier model, this is explained by a reduction of the electron mobility $\mu$ in that way, that it overcompensates the increase of $n$, see Eq. (S3).

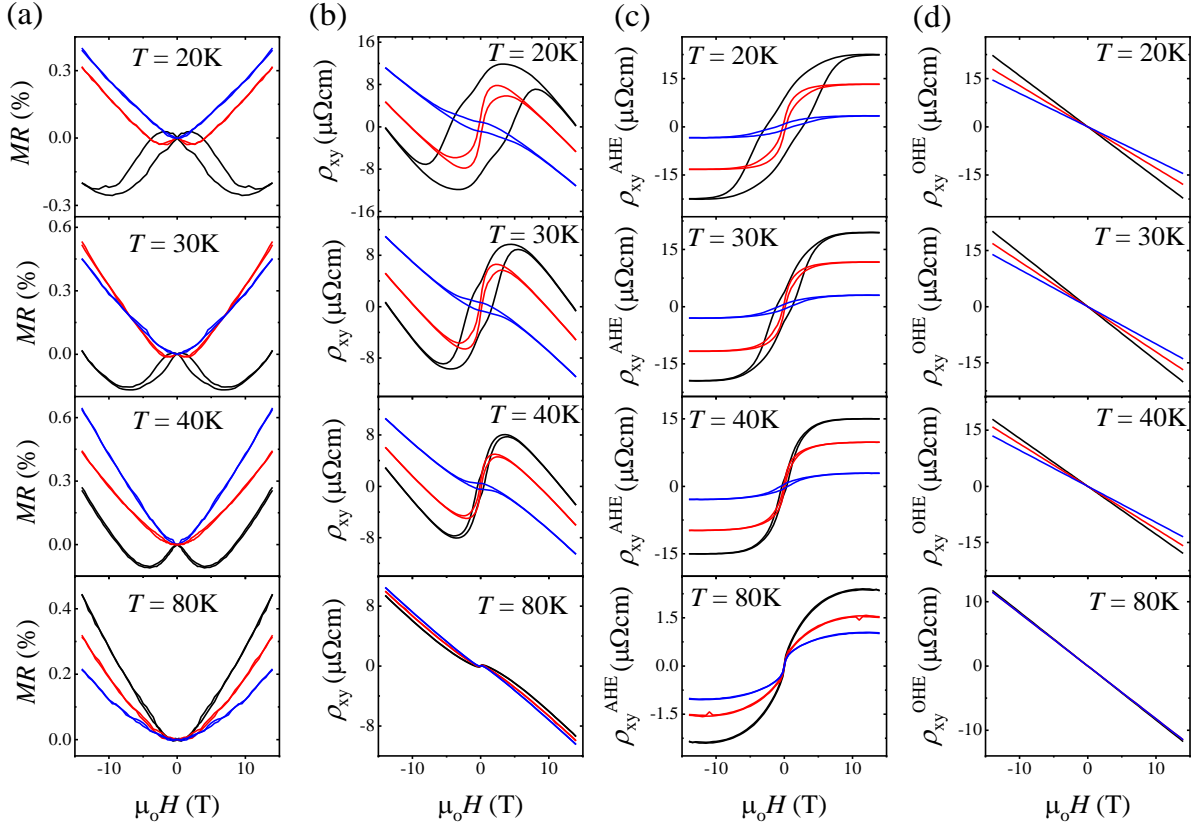

**Figure S6. Electric-field dependence of the magnetotransport for $T < T_C$.** The magnetoresistance $MR$ (**a**), Hall-resistivity $\rho_{xy}$ (**b**), the anomalous part $\rho_{AHE}$ of the Hall-resistivity $\rho_{xy}$ (**c**), and the ordinary part $\rho_{OHE}$ of $\rho_{xy}$ versus $\mu_0 H$ for different $V_g$ (blue: -50V, red: 0V, and black: +50V) for $T$ = 20, 30, 40, and 80K.

**Anomalous magnetoresistance.** The proximity induced magnetism of SIO results in an angle-dependent anisotropic magnetoresistance $AMR(\alpha) = [\rho_{xx}(\alpha)-\rho_{xx}(90°)]/\rho_{xx}(90°)$ below $T_C$, where $\alpha$ is the angle between the in-plane magnetic field- and current-direction. $AMR(\alpha)$ is well described by a two- and a four-fold component, *i. e.*, $AMR(\alpha) = C_0 + C_2(\alpha) + C_4(\alpha)$, where $C_2(\alpha) = <C_2> \times \cos(2\alpha - \omega_2)$ and $C_4(\alpha) = <C_4> \times \cos(4\alpha - \omega_4)$, with the amplitudes $C_0$, $<C_2>$, and $<C_4>$ and the offset angles $\omega_2$ and $\omega_4$. Figures S7a and S8a display $AMR(\alpha)$ for different $V_g$ at 5 and 10K for $H$ = 14 T. The corresponding two-fold and four-fold components are shown in the figures (b) and (c), respectively.

$C_2(\alpha)$ shows maxima at $\alpha = 0°$ (current parallel to $H$ and the $<100>_{pc}$ pseudo-cubic crystallographic direction) and 180°, whereas $C_4(\alpha)$ displays minima at $\alpha = 45°+n\times 90°$ ($n$ = 0, 1, 2, and 3). The amplitude $<C_2>$ increases strongly when $V_g$ changes from -50 V to 50 V. The increase is well comparable to the increase of $\rho_{AHE}^s$ and hence obviously related to the increase of the magnetization. In comparison, $<C_4>$ changes less with $V_g$ however still indicate enhanced

<110>-easy-axis behavior. The positions of the extrema are obviously not affected by $V_g$. For $T > 40$ K $AMR(\alpha)$ diminishes very much which makes distinct data analysis difficult.

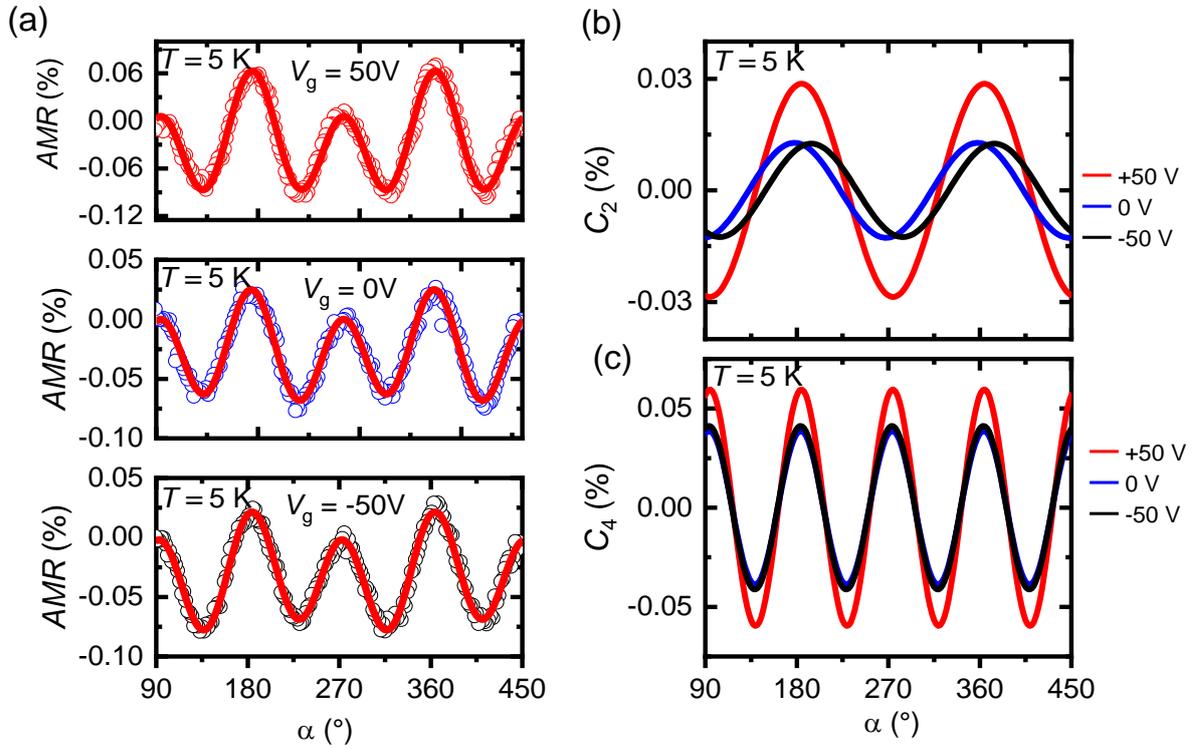

**Figure S7. Anomalous magnetoresistance of SIO/LCO heterostructures at $T$ = 5K.** (**a**) The angle-dependent anisotropic magnetoresistance $AMR(\alpha)$ at $T$ = 5K for $V_g$ = 50V, 0, and -50 V. (**b**) The two-fold component $C_2$ and (**c**) the four-fold component $C_4$ of $AMR(\alpha)$.

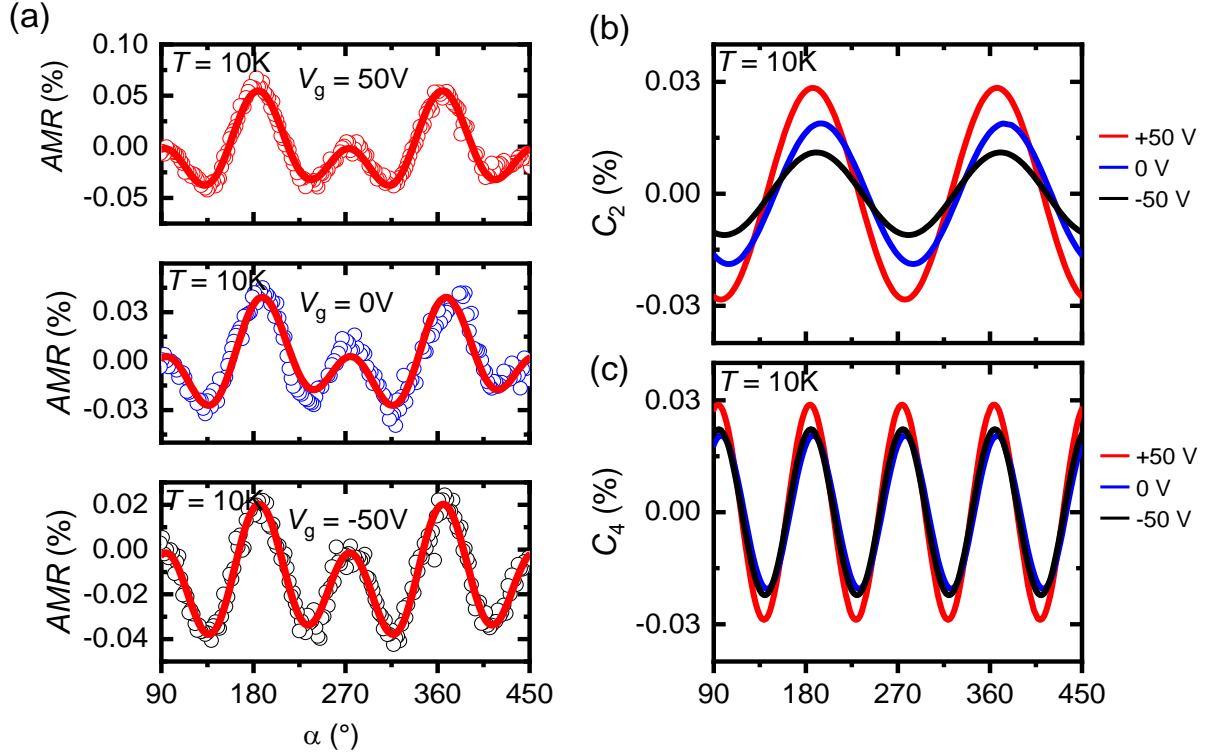

**Figure S8. Anomalous magnetoresistance of SIO/LCO heterostructures at $T$ = 10K.** (**a**) The angle-dependent anisotropic magnetoresistance $AMR(\alpha)$ at $T$ = 5K for $V_g$ = 50V, 0, and -50 V. (**b**) The two-fold component $C_2$ and (**c**) the four-fold component $C_4$ of $AMR(\alpha)$.

**Fitting parameters.** In the tables S1-S3 we have summarized all fitting parameters as deduced from fittings shown in the main manuscript, Fig. 2b and c, and Fig. 4a.

**Table S1**. Fitting parameters for fitting $\rho_{AHE}$ at $T$ = 20 K as shown in Fig. 2b. For the fitting, two Heaviside-step functions corresponding to $\rho_{xy}^{AHE} = a1 \tanh\left(\omega 1(B - B_{c,1})\right) + a2 \tanh\left(\omega 2(B - B_{c,2})\right)$ were used:

| V (V) | $a_1$ (μΩcm) | $\omega_1$ (T⁻¹) | $B_{c,1}$ (T) | $a_2$ (μΩcm) | $\omega_2$ (T⁻¹) | $B_{c,2}$ (T) |
|---|---|---|---|---|---|---|
| +50 | 16.41 | 0.196 | 0.816 | 6.31 | 0.503 | 5.573 |
| +30 | 15.78 | 0.263 | 0.526 | 4.04 | 0.631 | 2.547 |
| 0 | 9.31 | 0.647 | 0.116 | 3.99 | 0.199 | 1.532 |
| -30 | 0.65 | 1.459 | 2.22045E-14 | 4.31 | 0.210 | 1.107 |
| -50 | 1.60 | 0.227 | 2.22045E-14 | 1.83 | 0.222 | 2.446 |

**Table S2**. Fitting parameters for fitting *MR* at *T* = 20 K as shown in Fig. 2c. For the fitting, two Heaviside-step functions for the description of *M* were used, *i. e.*, $\text{MR}(\%) = aB^2 - bM^2 = aB^2 - b\left(M_{s,1} \tanh(\omega 1'(B - B1'_c)) + M_{s,2} \tanh(\omega 2'(B - B2'_c))\right)^2$:

| V (V) | a (T$^{-2}$) | b (T$^{-1}$) | $M_{s,1}$ (T$^{-1}$) | $\omega_1'$ (T$^{-1}$) | $B_1'_c$ (T) | $M_{s,2}$ (T$^{-1}$) | $\omega_2'$ (T$^{-1}$) | $B_2'_c$ (T) |
|---|---|---|---|---|---|---|---|---|
| +50 | 1.1123E-5 | 1.43608E-6 | 42.65 | 0.133 | 0.032 | 14.87 | 0.239 | 4.533 |
| +30 | 1.22407E-5 | 2.22099E-6 | 25.68 | 0.167 | 0.257 | 12.98 | 0.498 | 1.693 |
| 0 | 1.86491E-5 | 3.56194E-7 | 25.00 | 0.501 | 0.205 | 2.22045E-14 | 2.22045E-14 | 2.22045E-14 |

**Table S3**. Fitting parameters for fitting $AMR(\alpha) = [\rho_{xx}(\alpha)-\rho_{xx}(90°)]/\rho_{xx}(90°)$. A two- and four-fold component was used to fit the data corresponding to $AMR(\alpha) = C_0 + <C_2> \times \cos(2\alpha - \omega_2) + <C_4> \times \cos(4\alpha - \omega_4)$:

| T (K) | $V_g$ (V) | $C_0$ (%) | $C_2$ (%) | $\omega_2$ (°) | $C_4$ (%) | $\omega_4$ (°) |
|---|---|---|---|---|---|---|
| 20 | +50 V | 0.01337 | 0.03719 | 186.31 | 0.02176 | 2.27 |
| 20 | 0 V | -0.01844 | 0.0111 | 185.48 | 0.01961 | 2.33 |
| 20 | -50 V | -0.01283 | 0.00875 | 190.14 | 0.01644 | -10.14 |

References:

[S1] J. Terzic et al., Phys. Rev. B 91, 235147 (2015).